\documentclass[letter,twocolumn]{jpsj3}
\usepackage{color}
\usepackage{ulem}
\title{Partial 
Disorder 
in the Periodic Anderson Model on a Triangular Lattice} 

\author{Satoru HAYAMI\thanks{E-mail address: hayami@aion.t.u-tokyo.ac.jp}, Masafumi UDAGAWA, and Yukitoshi MOTOME 
}
\inst{Department of Applied Physics, University of Tokyo, 7-3-1 Hongo, Bunkyo-ku, Tokyo 113-8656, Japan 
}
\abst{
We report our theoretical results on the emergence of a partially-disordered state at zero temperature and its detailed nature in the periodic Anderson model on a triangular lattice at half filling. 
The partially-disordered state is characterized by coexistence of a collinear antiferromagnetic order on an unfrustrated honeycomb subnetwork and nonmagnetic state at the remaining sites.  
This state appears with opening a charge gap between a noncollinear antiferromagnetic metal and Kondo insulator while changing the hybridization and Coulomb repulsion. 
We also find a characteristic crossover in the low-energy excitation spectrum as a result of coexistence of magnetic order and nonmagnetic sites. 
The result demonstrates that the partially-disordered state is observed distinctly even in the absence of spin anisotropy, in marked contrast to the partial Kondo screening state found in the previous study for the Kondo lattice model.
}

\kword{periodic Anderson model, geometrical frustration, triangular lattice, partial disorder, charge disproportionation, quantum critical point, partial Kondo screening}

\begin{document}
\maketitle

Rare-earth compounds have drawn considerable interest, since they exhibit fascinating behaviors, such as heavy quasi-particle formation, superconductivity, and non-Fermi liquid behavior, as a consequence of the interplay between localized electrons and conduction electrons~\cite{Hewson}.
The major outcomes of this interplay can be understood in terms of the two competing interactions, the Ruderman-Kittel-Kasuya-Yosida (RKKY) interaction and the Kondo coupling: 
The former favors magnetic ordering~\cite{Ruderman,Kasuya,Yosida(1957)}, whereas the latter promotes singlet formation between conduction and localized electrons~\cite{Kondo,Yosida(1966)}.
Their competition, in general, leads to a quantum critical behavior between a magnetically ordered state and a Fermi liquid state~\cite{Doniach}, resulting in the rich phenomena in these systems. 

In the present study, we investigate the effect of geometrical frustration in the quantum critical region, with emphasis on the emergence of unusual orders. 
Experimentally, interesting coexistence of nonmagnetic Kondo singlet and magnetic order was observed in several rare-earth compounds with frustrated lattice structures, such as CePdAl~\cite{Donni,Oyamada} and UNi$_{4}$B~\cite{Mentink,Oyamada2}.
Such coexistence was theoretically studied by the mean-field approximation (MFA) of a classical pseudomoment model.~\cite{Lacroix,Dolores} 
More recently, two of the authors and their collaborators showed a possibility of such coexistence, called the partial Kondo screening (PKS), by sophisticated variational Monte Carlo calculations for the Kondo lattice model (KLM) as well as the Kondo necklace model~\cite{Motome}. 

While the existence of PKS is shown by explicitly taking into account quantum fluctuations and the spin-charge interplay~\cite{Motome}, its origin has not been fully understood.
In particular, it remains unclear whether the PKS appears in the absence of spin anisotropy. 
Moreover, the detailed nature of PKS, such as the electronic structure, was not clarified yet. 
The nature of phase transitions from the surrounding phases to the PKS state is also unclear. 
It is highly desirable to carry out the further analysis of the details of PKS phenomena. 

In this Letter, we investigate the ground state of a periodic Anderson model (PAM) on a triangular lattice. 
PAM is a parent model for KLM, and it includes charge degree of freedom in the localized level explicitly. 
The model allows a simple and straightforward MFA, which enables wide-ranging analysis compared to the previous numerical study. 
We will demonstrate that this itinerant electron model exhibits a partially-disordered (PD) state, a coexistence of a collinear antiferromagnetic (AF) order and nonmagnetic sites, without aid from spin anisotropy.
The detailed nature of the PD phase and the associated phase transitions will be presented. 

We consider PAM whose Hamiltonian is given by~\cite{Anderson} 
\begin{eqnarray}
{\mathcal{H} }  =\!\!\!\!\! &- \!\!\!\!\! &t \sum_{\langle i,j\rangle,\sigma}  
( c^{\dagger}_{i \sigma} c_{j \sigma}  + {\mathrm{H.c.}} ) 
- V\sum_{i ,\sigma} 
( c^{\dagger}_{i \sigma}f_{i \sigma}+{\mathrm{H.c.}} ) 
\nonumber \\
\!\!\!\!\! &+ \!\!\!\!\! &U \sum_{i} f^{\dagger}_{i \uparrow}f_{i \uparrow} f^{\dagger}_{i \downarrow} f_{i \downarrow}
+ E_{0} \sum_{i, \sigma} f^{\dagger} _{i \sigma} f_{i \sigma} ,
\label{Ham}
\end{eqnarray}
where $c_{i \sigma}^{\dagger}$($c_{i \sigma}$) and $f_{i\sigma}^{\dagger}$($f_{i \sigma}$) are the creation (annihilation) operators of conduction and localized electrons at site $i$ and spin $\sigma$, respectively. The sum of $\langle i,j\rangle$ is taken over the nearest-neighbor sites on the triangular lattice.
The first term in eq.~(\ref{Ham}) represents the kinetic energy of conduction electrons, the second term the onsite $c$-$f$ hybridization between conduction and localized electrons, the third term the onsite Coulomb interaction for localized electrons, and the fourth term the atomic energy of localized electrons. 
PAM delivers KLM in the large $U$ limit with one $f$ electron per site. 
Hereafter, we take $t=1$ as an energy unit and $V>0$, and fix the electron density at half filling; 
$n = \sum_{i,\sigma} \langle c_{i\sigma}^{\dagger} c_{i\sigma} + f_{i\sigma}^{\dagger} f_{i\sigma} \rangle / N= 2$, where $N$ is the total number of sites. 
We choose $E_0=-1$, while the following results are qualitatively the same for $E_0 \lesssim 0$ within the noninteracting bandwidth.

\begin{figure}[t]
\begin{center}
\includegraphics[width=70mm]{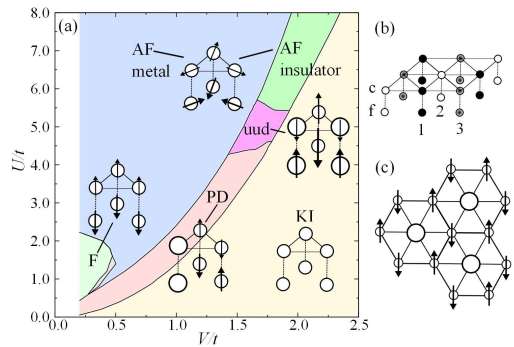}
\end{center}
\caption{(Color online). (a) Ground-state phase diagram of the model in eq.~(\ref{Ham}) at half-filling and $E_0=-1$. Schematic picture for the ordering pattern is shown for each phase. 
The size of the circles reflects the magnitude of local electron densities, and the arrows represent local spin moments. 
In the region for $V<0.2$ and the narrow grey region between the F and AF metal, it is difficult to obtain a converged mean-field solution.
(b) A three-site unit cell used in the Hartree-Fock calculation. 
Upper circles forming a triangular lattice and the dangling lower circles represent the $c$ and $f$ electron degrees of freedom, respectively. 
(c) Schematic picture of the PD state. One of the three sublattices becomes nonmagnetic, while the remaining two sublattices forming a honeycomb subnetwork retain a collinear AF order.
}
\label{phase diagram}
\end{figure}

In order to obtain the ground state phase diagram of the model (\ref{Ham}), we apply the standard Hartree-Fock approximation to decouple the Coulomb $U$ term with preserving SU(2) symmetry in spin space. 
We employ three-site unit cell, as shown in Fig.~\ref{phase diagram}(b), and calculate the mean fields by taking the sum over $200 \times 200$ grid points in the first Brillouin zone. 
The mean fields are determined self-consistently, by repeating calculations until they converge within the precision less than $10^{-6}$.

\begin{figure}[t]
\begin{center}
\includegraphics[height=75mm]{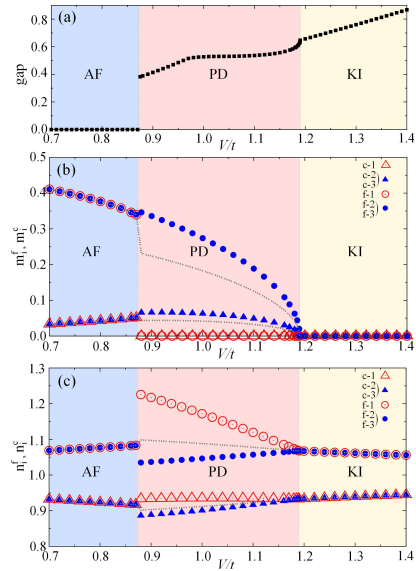}
\end{center}
\caption{(Color online) $V$ dependence of (a) energy gap, (b) magnitude of local spin moment and (c) local electron density at $U=2$ and $E_{0}=-1$. 
In (b) and (c), the data are plotted for $c$ and $f$ electrons at each sublattice separately, and the dashed curves represent the mean values of $c$ and $f$ contributions respectively.
Note that site 2 and 3 are equivalent in each phase.
}
\label{physical quantity}
\end{figure}

Figure~\ref{phase diagram}(a) shows the result for the ground-state phase diagram. 
The phase diagram is roughly divided into three regions; magnetic region for $U \gtrsim 2V^2$, nonmagnetic region for $U \lesssim 2V^2$, and intermediate region for $U \sim 2V^2$. 
Figure~\ref{physical quantity} shows the charge gap, magnitude of local spin $m_i^{f(c)} = \sqrt{\langle \mathbf{s}_{i,x}^{f(c)} \rangle^{2} + \langle \mathbf{s}_{i,y}^{f(c) } \rangle^{2} + \langle \mathbf{s}_{i,z}^{f(c)} \rangle^{2}}$ [$\mathbf{s}_{i,\mu}^{f(c)}$ is the $\mu$ component of the spin operator for $f$($c$) electron], and local electron density $n_i^{f(c)} = \langle \sum_\sigma f_{i\sigma}^\dagger f_{i\sigma}(c_{i\sigma}^\dagger c_{i\sigma}) \rangle$, for these three regions while varying $V$ at $U=2$.

In the magnetic region for $U \gtrsim 2V^2$, a $120^{\circ}$ AF order is dominantly stabilized. 
This state is a gapless metal, as shown in Fig.~\ref{physical quantity}(a).
The magnetic moments of conduction and localized electrons form 120$^{\circ}$ order in the entire region of the AF metal, as schematically shown in Fig.~\ref{phase diagram}(a) [see also Fig.~\ref{physical quantity}(b)], except for a slight canting from the 120$^{\circ}$ order seen for $U \gtrsim 6$ and $V \lesssim 0.4$. 
The magnetic order is likely to be understood from the strong-$U$ KLM picture, in which the RKKY interaction should favor AF configuration between neighboring spins. 
In the smaller $U$ region, the system becomes ferrimagnetic (F), whose magnetic state is schematically shown in Fig.~\ref{phase diagram}(a), presumably originating in the Stoner-like mechanism. 

On the other hand, in the regime for $U \lesssim 2V^2$, the system becomes insulating [Fig.~\ref{physical quantity}(a)] and nonmagnetic [Fig.~\ref{physical quantity}(b)].
There, the $c$-$f$ hybridization $V$ gives rise to bonding and anti-bonding states, and the bonding states are fully occupied at half filling.
This band insulator is called the Kondo insulator (KI). 

Between the magnetic and nonmagnetic regimes, the system exhibits a PD phase for $U \lesssim 4.5$. 
In this phase, one of the three sublattices becomes nonmagnetic, while the remaining two sublattices retain magnetic moments [Fig~\ref{physical quantity}(b)]; see below for the detail.  
For larger $U$, the system turns into an up-up-down type ferrimagnetic collinear state (uud) at $U\sim 5$, and finally a 120$^\circ$ AF insulator for $U \gtrsim 5.5$, as shown in Fig.~\ref{phase diagram}(a)~\cite{note_uud}. 

Let us examine the nature of the intermediate PD state and the associated phase transitions. 
First of all, this state is insulating; the charge gap is nonzero, as plotted in Fig.~\ref{physical quantity}(a). 
In the PD phase, the gap grows with increasing $V$, while its increase is suppressed for $V \gtrsim 0.98$ after showing a shoulder-like behavior. We will return to this peculiar dependence later. 
Our result shows that the transition to the 120$^\circ$ AF metallic state is of first order with a jump of the gap, while it is continuous when entering to the Kondo insulator. 
We note that the results, especially the first-order transition between the noncollinear and collinear states, might be modified when taking account of a longer period order beyond the three-site unit cell, especially in the first-order transition between the noncollinear and collinear orderings. 

Second, the PD state exhibits peculiar coexistence of magnetic and nonmagnetic sites; the magnetic moments of both conduction and localized electrons vanish on one of three sublattices, while they remain finite on the remaining two sublattices, as shown in Fig.~\ref{physical quantity}(b). 
The moment is larger for localized electrons than for conduction ones, and the magnitude of each moment takes the same value between the two sublattices, as in the 120$^\circ$ AF metallic state for smaller $V$. 
The magnetic moments, however, show a collinear AF order on the two sublattice sites [Fig.~\ref{phase diagram}(a)]. 
Therefore, the PD state is realized by relaxing the geometrical frustration; that is, the system becomes free from the magnetic frustration by spontaneously self-organizing into the unfrustrated honeycomb AF network and the remaining nonmagnetic sites, as schematically shown in Fig.~\ref{phase diagram}(c). 
Note that this peculiar state is obtained as the ground state, qualitatively different from 
the entropically-driven PD states sometimes found in frustrated classical spin systems~\cite{Mekata}. 

Finally, the PD state is accompanied by charge disproportionation: the local charge density is higher at the nonmagnetic site than the AF sites, as shown in Fig.~\ref{physical quantity}(c). 
Near the transition into the KI phase, the charge disproportionation decreases linearly with decreasing the distance from the critical point, in contrast to the square-root behavior of the local spin moment in Fig.~\ref{physical quantity}(b). 
This indicates that the collinear AF moment is a primary order parameter and the charge disproportionation is induced as a secondary effect. 

\begin{figure}[t]
\begin{center}
\includegraphics[width=70mm]{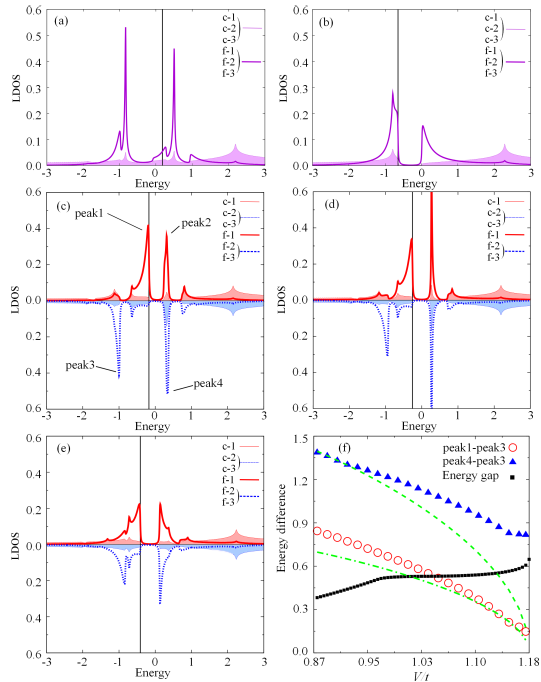}
\end{center}
\caption{(Color online). 
LDOS of conduction and localized electrons for each phase at $U=2$ and $ E_{0}=-1$. 
The data are taken at (a) $V=0.87$, (b) 1.20, (c) 0.90, (d) 0.98, and (e) 1.10.
(a) and (b) correspond to the 120$^\circ$ AF metallic phase and the KI phase, respectively, while (c)-(e) are for the PD phase. 
The thin vertical lines show the Fermi levels defined by the energy of the highest occupied level. 
Note that all the sites are equivalent in (a) and (b), and that the site 2 and 3 are equivalent in (c)-(e). 
(f) $V$ dependence of the energy differences between the peak 3 and 4 and between the peak 3 and 1. 
The dashed and dot-dashed curves denote $2 U m_{2(3)}^f$ and $U m_{2(3)}^f$, respectively. 
For comparison, the energy gap in Fig.~\ref{physical quantity}(a) is also plotted. 
}
\label{DOS}
\end{figure}

Now, we discuss the phase changes from the viewpoint of the electronic state. 
Figure~\ref{DOS} shows the local density of states (LDOS) for conduction and localized electrons for several $V$ at $U=2$. 
In the AF phase, the energy spectrum is modulated due to the presence of static AF magnetic moments. 
As a result, there appear two sharp peaks, as shown in Fig.~\ref{DOS}(a), whose energy separation is given by the mean-field Coulomb exchange energy $2U m_{i}^f$. 
On the other hand, in the KI phase, there appear two peaks on both sides of the energy gap, as shown in Fig.~\ref{DOS}(b). 
Their origin, however, is not magnetic but the $c$-$f$ hybridization $V$. 
The peaks consist of the hybridized bands, which have a major contribution from the localized $f$ levels. 

In the intermediate PD phase, the energy spectrum basically shows coexistence of the characteristic features of both the AF and KI phases [Figs.~\ref{DOS}(c)-(e)]. 
At the nonmagnetic site (upper half in the figures), the LDOS shows two prominent peaks indicated by the peak 1 and 2 in Fig.~\ref{DOS}(c). 
They appear on both sides of the gap and are dominated by the contribution from localized electrons. 
This is characteristic of the hybridization effect in Fig.~\ref{DOS}(b). 
Meanwhile, at the magnetic sites (lower half), the LDOS exhibits two sharp peaks (peak 3 and 4). 
The energy difference between the two peaks is close to $2U m_{2(3)}^f$ as in the AF state, at least, in the region near the AF metallic phase, where the collinear AF moment is well developed [see also Fig.~\ref{DOS}(f)]. 
In addition, the energy difference between the peak 3 and 1 is scaled approximately to $U m_{2(3)}^f$, as plotted in Fig.~\ref{DOS}(f). 
This indicates that the system gains energy to stabilize the PD state from the formation of the collinear AF order on the self-organized unfrustrated honeycomb network.

With approaching the transition to the KI by increasing $V$, the gap-edge contributions at the nonmagnetic site (peak 1 and 2) smoothly vary to the peaks of the hybridized bands in the KI [Figs.~\ref{DOS}(c)-(e)]. 
On the other hand, the LDOS at the AF sites changes its nature considerably with increasing $V$. 
As plotted in Fig.~\ref{DOS}(f), the energy difference of the two peaks scales well to $2U m_{2(3)} ^{f}$ for small $V$ as mentioned above, but it starts to deviate for $V \gtrsim 0.95$: Finally, the two peak structures merge into the hybridized bands in the KI phase. 
This behavior is considered to be a crossover from the AF dominant to hybridization dominant regime. 

\begin{figure}[t]
\begin{center}
\includegraphics[height=60mm]{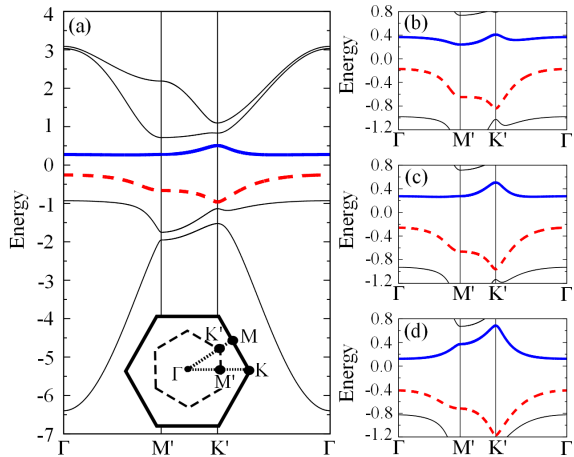}
\end{center}
\caption{
(Color online). 
Energy dispersion at $U=2$ and $E_{0}=-1$. 
The results are shown along the $\Gamma$-M'-K'-$\Gamma$ line in the Brillouin zone, as shown in the inset of (a). 
In the inset, the solid hexagon represents the first Brillouin zone and the dashed one is for the folded Brillouin zone under three-sublattice ordering. 
(a) Overall structure of the energy dispersion at $V=0.98$. 
(b)-(d) Energy dispersion in the vicinity of the Fermi level at (b) $V=0.90$, (c) $0.98$, and (d) $1.10$, respectively.
The bold(blue) and bold-dashed(red) lines denote LUB and HOB. 
}
\label{dispersion}
\end{figure}

We further analyze the qualitative change of energy spectrum in the PD state in terms of the band structure.
Figure~\ref{dispersion}(a) shows the typical band structure in the PD phase. 
There are six bands (doubly degenerate each) under the three-sublattice ordering. 
The overall structure does not change so much in the entire region of the PD phase, but an interesting qualitative change appears near the Fermi level while increasing $V$. 
As shown in Figs.~\ref{dispersion}(b)-(d), although the highest occupied band (HOB) does not show remarkable change, the lowest unoccupied band (LUB) exhibits nontrivial change. 
For small $V$ [Fig.~\ref{dispersion}(b)], the bottom of the LUB resides at the M' point in the Brillouin zone. 
This is attributed to the well-developed threefold AF order which opens a gap at the M' point. 
With increasing $V$, the energy at the $\Gamma$ point gradually decreases relatively to the M' point, and becomes lower for $V \gtrsim 0.98$ [Figs.~\ref{dispersion}(c) and \ref{dispersion}(d)]. 
Thus, the gap structure changes from the indirect one between $\Gamma$ and M' to the direct one at the $\Gamma$ point. 
The latter smoothly turns into the hybridization gap in the KI~\cite{note_KIgap}. 
Hence, the result shows a crossover of the LUB nature from the AF dominant to hybridization dominant.

The crossover of LUB is clearly reflected in the shoulder-like feature in the $V$ dependence of energy gap [Figs.~\ref{physical quantity}(a) and \ref{DOS}(f)]. 
For $V \lesssim 0.98$ where the LUB is minimum at the M' point, the indirect gap increases with $V$. Meanwhile, for $V \gtrsim 0.98$ where the LUB bottom comes at the $\Gamma$ point, the gap does not show substantial $V$ dependence, as the energy difference between HOB and LUB is almost constant at the $\Gamma$ point. 

The peculiar behavior of LUB implies rich possibilities caused by electron doping to the PD state.
In particular, in the crossover region, the LUB becomes extremely flat, as shown in Fig~\ref{dispersion}(c), which leads to a sharp peak of LDOS at the upper edge of the gap [Fig.~\ref{DOS}(d)]. 
This suggests a possibility to have coexistence of heavy-fermion state and magnetic ordering, as experimentally observed in CePdAl~\cite{Oyamada}. 
It will be interesting to investigate such carrier doping effect, possibly by more sophisticated method than the MFA. 

Finally, let us discuss our results in relation with the previous theoretical study for KLM~\cite{Motome}.
As mentioned above, KLM corresponds to the large $U$ limit of PAM with one $f$ electron per site; the $f$ electrons give localized moments, which couple with conduction electrons via the Kondo coupling $J \propto V^2/U$. 
Hence, the phase sequence of AF, PD, and KI while changing $V^2/U$ in our results for PAM appears to well correspond to that of AF, PKS, and Kondo spin liquid (KSL) while changing $J$ in KLM.  
The coincidence, however, needs careful consideration. 
The point is that our PD state takes place even in the absence of the spin anisotropy, whereas the PKS in KLM is hardly stabilized without spin anisotropy. 
Furthermore, our PD state appears in the relatively weak $U$ regime, far from the strong $U$ limit where KLM is justified.
These results indicate that, besides the spin anisotropy, the charge degree of freedom via the $c$-$f$ hybridization plays an important role in the stabilization of the PD state.  

When the spin anisotropy is introduced into the present PAM, the PD state is expected to extend to a wider range of parameters, and is presumably connected to the PKS state found in the strong-$U$ limit, i.e., in KLM. 
Furthermore, we expect that the PD state in PAM is further stabilized by taking quantum fluctuations into account beyond the MFA, since the previous study indicated that quantum fluctuations stabilize PKS~\cite{Motome}. 
Such extensions are left for future study.  

In summary, we have investigated the ground state of the periodic Anderson model at half filling on a triangular lattice by the Hartree-Fock approximation.
We found that the model exhibits a partial disorder with coexistence of magnetic order and nonmagnetic sites, which was not apparently observed in the related Kondo lattice model in the absence of spin anisotropy. 
We clarified the phase diagram in a wide parameter range and the nature of phase transitions from the surrounding phases to the partially-disordered phase. 
We also revealed the detailed nature of the partially-disordered state: it appears with gap opening and shows a characteristic crossover behavior in the structure of the excitation spectrum from the dominantly-antiferromagnetic to dominantly-Kondo-insulating regime. 

\noindent {\bf Acknowledgements}\\
We acknowledge helpful discussions with H. Uchigaito and J. Yoshitake.
This work was supported by KAKENHI (No. 19052008, No. 21340090, and No. 21740242), Global COE Program ``the Physical Sciences Frontier", from the Ministry of Education, Culture, Sports, Science and Technology, Japan.


\begin{thebibliography}{99}
\bibitem{Hewson} A. C. Hewson: {\it The Kondo Problem to Heavy Fermions} (Cambridge University Press, Cambridge, England, 1993), and references therein.
\bibitem{Ruderman} M. A. Ruderman and C. Kittel: Phys. Rev. {\bf 96} (1954) 99.
\label{Ruderman}
\bibitem{Kasuya} T. Kasuya: Prog. Theor. Phys. {\bf 16} (1956) 45.
\label{Kasuya}
\bibitem{Yosida(1957)} K. Yosida: Phys. Rev. {\bf 106} (1957) 893.
\label{Yosida(1957)}
\bibitem{Kondo} J. Kondo: Prog. Theor. Phys. {\bf 32} (1964) 37.
\label{Kondo}
\bibitem{Yosida(1966)} K. Yosida: Phys. Rev. {\bf 147} (1966) 223.
\label{Yosida(1966}
\bibitem{Doniach} S. Doniach: Physica 
{\bf 91B} (1977) 231.
\label{Doniach}
\bibitem{Donni} A. D\"{o}nni, G. Ehlers, H. Maletta, P. Fischer, H. Kitazawa, and M. Zolliker: J. Phys.: Condens. Matter {\bf 8} (1996) 11213.
\label{Donni}
\bibitem{Oyamada} A. Oyamada, S. Maegawa, M. Nishiyama, H. Kitazawa, and Y. Isikawa: Phys. Rev. B {\bf 77} (2008) 064432.
\label{Oyamada}
\bibitem{Mentink} S. A. M. Mentink, A. Drost, G. J. Nieuwenhuys, E. Frikkee, A. A. Menovsky, and J. A. Mydosh: Phys. Rev. Lett. {\bf 73} (1994) 1031.
\label{Mentink}
\bibitem{Oyamada2} A. Oyamada, M. Kondo, K. Fukuoka, T. Itou, S. Maegawa, D. X. Li, and Y. Haga: J. Phys.: Condens. Matter {\bf 19} (2007) 145246.
\bibitem{Lacroix} C. Lacroix, B. Canals, and M. D. N\'u$\tilde{\rm n}$ez-Regueiro: Phys. Rev. Lett. {\bf 77} (1996) 5126. 
\bibitem{Dolores} M. D. N\'u$\tilde{\rm n}$ez-Regueiro, C. Lacroix, and B. Canals: Physica C {\bf 282}-{\bf 287} (1997) 1885. 
\bibitem{Motome} Y. Motome, K. Nakamikawa, Y. Yamaji, and M. Udagawa: Phys. Rev. Lett. {\bf 105} (2010) 036403.
\label{Motome}
\bibitem{Anderson} P. W. Anderson: Phys. Rev. {\bf 124} (1961) 41.
\label{Anderson}
\bibitem{note_uud} The result in the intermediate to large $U$ region might be 
modified when going beyond 
the Hartree-Fock approximation. 
\bibitem{Mekata} M. Mekata: J. Phys. Soc. Jpn. {\bf 42} (1977) 76.
\label{Mekata}
\bibitem{note_KIgap} The hybridization gap in KI is an indirect one between 
$\Gamma$ and K, which are equivalent in the folded Brillouin zone. 
\end{thebibliography}
\end{document}